\documentclass{wiley-article}

\usepackage[super]{natbib}
\makeatletter
\renewcommand{\@biblabel}[1]{#1.}
\makeatother

\usepackage{siunitx}
\usepackage{amssymb}
\usepackage{multirow}
\definecolor{hyperref-blue}{RGB}{0,0,160}
\definecolor{symbol-red}{RGB}{192,0,0}
\usepackage[colorlinks,linkcolor=hyperref-blue, anchorcolor=hyperref-blue, citecolor=hyperref-blue]{hyperref}
\usepackage[ruled,linesnumbered]{algorithm2e}
\papertype{RESEARCH ARTICLE}
\paperfield{International Journal of Intelligent Systems}

\title{Exploring Lottery Ticket Hypothesis in Media Recommender Systems}

\author[1]{Yanfang Wang}
\author[1]{Yongduo Sui}
\author[2]{Xiang Wang}
\author[3]{Zhenguang Liu}
\author[1\authfn{1}]{Xiangnan He}

\affil[1]{School of Information Science and Technology, University of Science and Technology of China, Hefei, Anhui, 230022, China}
\affil[2]{School of Computing, National University of Singapore, Kent Ridge, 119077, Singapore}
\affil[3]{College of Computer Science and Technology, Zhejiang University,  Hangzhou, Zhejiang, 310007, China}

\corraddress{Xiangnan He, School of Information Science and Technology, University of Science and Technology of China, Hefei, Anhui, 230022, China}
\corremail{hexn@ustc.edu.cn}

\fundinginfo{National Key Research and Development Program of China, Grant/Award Number: 2020YFB1406703; National Natural Science Foundation of China, Grant/Award Number: U19A2079, 61972372 and 62121002}

\conflict{The authors declared that they have no conflicts of interest to this work.}
\contribution{All persons who meet authorship criteria are listed as authors, and all authors certify that they have participated sufficiently in the work to take public responsibility for the content.}

\runningauthor{Wang~\textit{et al.}}

\begin{document}

\begin{frontmatter}
\maketitle

\begin{abstract}
Media recommender systems aim to capture users' preferences and provide precise personalized recommendation of media content. There are two critical components in the common paradigm of modern recommender models: (1) representation learning, which generates an embedding for each user and item; and (2) interaction modeling, which fits user preferences towards items based on their representations. In spite of great success, when a great amount of users and items exist, it usually needs to create, store, and optimize a huge embedding table, where the scale of model parameters easily reach millions or even larger. Hence, it naturally raises questions about the heavy recommender models: Do we really need such large-scale parameters? We get inspirations from the recently proposed lottery ticket hypothesis (LTH), which argues that the dense and over-parameterized model contains a much smaller and sparser sub-model that can reach comparable performance to the full model. In this paper, we extend LTH to media recommender systems, aiming to find the winning tickets in deep recommender models. To the best of our knowledge, this is the first work to study LTH in media recommender systems. With MF and LightGCN as the backbone models, we found that there widely exist winning tickets in recommender models. On three media convergence datasets --- Yelp2018, TikTok and Kwai, the winning tickets can achieve comparable recommendation performance with only 29\% \textasciitilde 48\%, 7\% \textasciitilde 10\% and 3\% \textasciitilde 17\% of parameters, respectively.

\end{abstract}
\end{frontmatter}

\section{Introduction}
In the era of information explosion, media recommendation is becoming the core of many online platforms, to accurately yield the information of interest that meets user needs. Media recommender systems aim to provide precise personalized recommendation for users to find their preferences from the deluge of items. Towards better media recommendation, extensive deep recommender models\cite{liu2021an,chen2021trust,wang2019neural,he2020lightgcn} have been proposed to capture user preference from behavioral data.
In general, the common paradigm of these models can be systematized as two critical components:
(1) representation learning, which creates a representation vector (i.e., embedding) for each single user and item;
and (2) interaction modeling, which fits the historical user-item interactions and predicts the preference of a user to an item based on their embeddings~\cite{batmaz2019review, zhang2019deep}. Obviously, the capacity of recommender model depends heavily on the representation ability of user and item embeddings, which are influenced by the scale of parameters --- an embedding with the larger size not only delineates the characteristics of users and items better~\cite{he2021automl}, but also predicts the user-item interaction more precisely~\cite{gupta2020deeprecsys}. Although the large embedding table brings obvious accuracy improvements, it suffers from two main limitations:

(1) The massive parameters become the major obstacle for production deployment and real-time prediction of deep recommender models. Specifically, in real-world scenarios, the number of users and items is usually up to tens of millions or even hundreds of millions. If each user (or item) is associated with a certain dense embedding, it will bring extremely expensive costs of memory and inference time to maintain the huge embedding table for all users and items. A good case is the famous audio-visual media convergence recommender system YouTube Recommendation~\cite{covington2016deep}, which projects a million videos into a 256-dimensional vector space, producing an embedding table with 256 million model parameters.

(2) Interaction histories of different users usually contain varying item numbers and sets, thus having different informativeness. Nevertheless, current recommender models almost assign all users and items with the same embedding size.
Hence, unifying the embedding size might limit the representation ability of user and item embeddings. Especially in a large-scale industrial media recommendation scenario, it may cause serious over-parameterization and over-fitting issues.

One prevalent solution is to adopt unstructured pruning (a.k.a. sparse pruning)~\cite{liu2020learnable} on the user-item embedding table. As a branch of neural network pruning technology, sparse pruning aims to refine sparse sub-networks with similar performance but fewer parameters by removing redundant parameters of dense networks. In traditional sparse pruning, a large over-parameterized dense network is required to be trained first, followed by pruning and fine-tuning. Obviously, training a large dense network is expensive. Therefore, some researchers~\cite{frankle2018lottery} suggested that, rather than training a large dense network and then pruning it into a small sparse sub-network, it is better to train a small sparse sub-network from the scratch. This view is normally considered infeasible in previous studies, because large dense networks have wide parameter search spaces to find optimum solution through independent training, while small sparse networks merely have limited parameter search spaces.

More recently, Frankle and Carbin~\cite{frankle2018lottery} proposed Lottery Ticket Hypothesis (LTH): a large and dense deep neural network contains a small sparse sub-network, which can match the test accuracy of the original large network for at most the same number of iterations when trained independently. This trainable sparse sub-network is called a Winning Ticket. At present, LTH has been verified in the fields of Computer Vision (CV)~\cite{frankle2018lottery} and Natural Language Processing (NLP)~\cite{yu2019playing}. However, to the best of our knowledge, there is no relevant research in the field of recommendation. If we can verify the existence of winning tickets and find them in deep recommender models, then the model parameters can be effectively reduced while preserving the test performance. Benefiting from the technique of \textit{sparse matrix storage}~\cite{virtanen2020scipy}, we can not only reduce the memory usage, but also speed up the inference.

In this work, we explore LTH in media recommender systems.
Specifically, we focus on the widely-used model parameters --- the user-item embedding table. Following the previous studies~\cite{frankle2018lottery}, we speculate that a large dense user-item embedding table contains a smaller sparser sub-matrix, which can be independently trained to achieve a similar performance as the full matrix. Technically, we exploit the iterative magnitude-based pruning (IMP)~\cite{frankle2018lottery,chen2021unified}, which gradually creates the embedding's binary masks based on their weight magnitudes and eventually reaches a powerful sub-matrix. The sub-matrix is referred to as a winning ticket in the recommender model. We conduct the extensive experiments on three benchmark datasets, where the empirical results consistently show that: (1) \textbf{winning tickets exist widely in deep recommender models}; (2) \textbf{the IMP algorithm can stably find the winning tickets}. Meanwhile, the winning tickets we found significantly outperform the original large matrix in terms of time consumption, memory usage and test performance.

Our main contributions are as follows:
\begin{itemize}
       \item We explore lottery ticket hypothesis in media recommender systems, aiming to reduce the memory usage and speed up the inference while preserving the recommendation accuracy. To the best of our knowledge, this is the first work to study lottery ticket hypothesis in the field of recommendation.
       \item With two deep recommender models Matrix Factorization (MF)~\cite{rendle2009bpr} and Light Graph Convolution Networks (LightGCN)~\cite{he2020lightgcn} as the backbone models being pruned, we conduct a lot of experiments on three real-world media convergence recommendation datasets, demonstrating that winning tickets widely exist in deep recommender models and the IMP algorithm we used can stably find the winning tickets.
       \item The experimental results show that the winning tickets are significantly superior over the original large embedding table in terms of time consumption, memory usage, and test performance. Specifically, on three datasets Yelp2018, TikTok and Kwai, our found winning tickets achieve the same performance as the complete models with only 29\% \textasciitilde 48\%, 7\% \textasciitilde 10\% and 3\% \textasciitilde 17\% of the parameters, respectively.
\end{itemize}

\section{Related work}

\subsection{Lightweight Embedding}

As the most basic parameters in the deep recommender models, the huge user-item embedding table dominates both the parameter scale and the inductive bias of the model. In order to choose an appropriate embedding size and reduce the parameter scale of the embedding table, there are two promising lightweight techniques:

\textbf{Auto Machine Learning (AutoML).}\cite{he2021automl} Joglekar~\textit{et al.}~\cite{joglekar2020neural} designed two neural search approaches: Single-Size embedding (NIS-SE) and Multi-Size embedding (NIS-ME). They defined the embedding table search space by blocks and adjusted the controller with the validation set, making the model automatically determine the embedding size by maximizing the accuracy under the constraint of the embedding table memory. On top of that, some other neural search approaches such as Neural Architecture Search (NAS)~\cite{liu2020automated}, Efficient Neural Architecture Search (ENAS)~\cite{pham2018efficient}, and Differentiable Architecture Search (DAS)~\cite{liu2018darts} are also utilized to automatically determine the size of embeddings and some of them have been deployed in the industry media recommender systems and achieved certain benefits.

\textbf{Model Compression.}~\cite{lei2018survey,cheng2018model} Common model compression technologies mainly includes pruning~\cite{xiang2021one}, quantization~\cite{gupta2015deep} and distillation~\cite{jiao2020tinybert}. For example, Liu~\textit{et al.}~\cite{liu2020learnable} compressed the embedding table by pruning the embedded vectors of each feature domain of the data. Sun~\textit{et al.}~\cite{sun2020generic} proposed a general sequential model compression framework, which decomposes the embedding table into multiple low-rank matrices. Wu~\textit{et al.}~\cite{wu2020saec} clustered the features for different domains of users and items by similarity calculation, effectively reducing the total feature number. Zhang~\textit{et al.}~\cite{zhang2020model} shared the parameters among similar features so as to reduce the parameters of embedding tables.

\subsection{Lottery Ticket Hypothesis}

For preserving accuracy, conventional sparse pruning methods have to train a large dense neural network first, and then prune it into a small sparse sub-network. Obviously, this process is computationally expensive. To address this problem, Frankle and Carbin~\cite{frankle2018lottery} proposed lottery ticket hypothesis, which states that there exist some small sparse sub-networks (winning tickets), which can be directly trained from the scratch to achieve performance comparable to the complete model with a similar or even faster training speed. To find the winning tickets in deep neural networks, they further proposed the IMP algorithm.

Recently, many efforts have been devoted to improving or extending lottery ticket hypothesis. Brix~\textit{et al.}~\cite{brix2020successfully} applied LTH to Transformer models, by rewinding the pruned sub-network weights to the values at iteration k instead of 0. Renda~\textit{et al.}~\cite{renda2020comparing} compared three different retraining techniques: fine-tuning, weight rewinding and learning rate rewinding. In their experiments, the IMP approach with weight rewinding achieved the best performance in terms of accuracy, compression ratio and search efficiency. You~\textit{et al.}~\cite{you2020drawing} modified the IMP algorithm to improve the search efficiency for winning tickets. Morcos~\textit{et al.}\cite{morcos2019one} generalized LTH across different datasets and optimizers, while Yu~\textit{et al.}~\cite{yu2019playing} extended it to the field of NLP and reinforcement learning, confirming the generality of winning tickets.

\section{Methodology}

This section begins with notations and definitions of the recommendation task (\ref{sec3.1}), followed by an introduction of the two widely-used deep recommender models MF and LightGCN (\ref{sec3.2}). Then we formally define LTH-MRS (\ref{sec3.3}), and detail our used method for searching winning tickets in recommender models (\ref{sec3.4}). Finally, we will provide some complexity analyses to show the superiority of the winning ticket (\ref{sec3.5}).

\subsection{Notations and Definitions}\label{sec3.1}

For a media recommender system containing $M$ users and $N$ items, let $\mathcal{U}=\{u_1, \cdots, u_M\}$ denotes the user set and $\mathcal{I}=\{i_1, \cdots, i_N\}$ denotes the item set. Let the node set $\mathcal{V}=\{v_1, \cdots, v_{|\mathcal{V}|}\}$ denotes all the users and items, where $\mathcal{V} = \mathcal{U} \cup \mathcal{I}$ and $|\mathcal{V}|=M+N$. Let the edge set ${\cal E}{\rm{ = }}\left\{ {\left( {u,i} \right)|u \in {\cal U},i \in {\cal I},{\mathop{\rm rec}\nolimits} (u,i) = 1} \right\}$ denotes the interactions between users and items, where ${\mathop{\rm rec}\nolimits} (u,i) = 1$ if there exists an interaction between the user $u$ and the item $i$, and $0$ otherwise. Then we can define the user-item interaction records as a bipartite graph ${\cal G} = \left\{ {{\cal V},{\cal E}} \right\}$. Learning with ${\cal G}$, the purpose of deep recommender systems is to predict the item preference ranking of each user.

For each node $v_t$ of ${\cal G}$, we represent it as an $F$-dimentional vector ${\mathbf{x}_t} \in {\mathbb{R}^F}$. Then the node embedding table of the whole graph ${\cal G}$ can be represented as follows:
\begin{equation}
       {\mathbf{X}}{\text{ = \{ }}\underbrace {{{\mathbf{x}}_1},{{\mathbf{x}}_2}, \cdots ,{{\mathbf{x}}_M}}_{{\text{users}}};{\text{ }}\underbrace {{{\mathbf{x}}_{M + 1}},{{\mathbf{x}}_{M + 2}}, \cdots ,{{\mathbf{x}}_{M + N}}}_{{\text{items}}}{\text{\} }} \in {\mathbb{R}^{\left| \mathcal{V} \right| \times F}}.
\end{equation}

We define the user-item interaction matrix as $\mathbf{R} \in {\mathbb{R}^{M \times N}}$, where
\begin{equation}
       \mathbf{R}[i,j] = \begin{cases}
              1 \quad \text{ if } (v_i, v_j) \in \mathcal{E}, \\
              0 \quad \text{otherwise}.
          \end{cases}       
\end{equation}

Then, we can obtain the adjacency matrix of ${\cal G}$ as follows:
\begin{equation}
       \mathbf{A} = {\text{ }}\left( {\begin{array}{*{20}{c}}
              0&{\mathbf{R}} \\ 
              {{{\mathbf{R}}^\mathrm{T}}}&0 
            \end{array}} \right) \in \mathbb{R}^{\left| \mathcal{V} \right| \times \left| \mathcal{V} \right|}.
\end{equation}

\subsection{Backbone Model}\label{sec3.2}
Taking the classical MF~\cite{rendle2009bpr} and the advanced LightGCN~\cite{he2020lightgcn} as the backbone models, we attempt to explore lottery ticket hypothesis in media recommender systems.

\subsubsection{MF}

MF~\cite{rendle2009bpr} is a classic recommender model, which aims to optimize the user-item embedding table to fit the user-item interaction records. We define the objective function of MF with bayesian personalized ranking (BPR)~\cite{rendle2009bpr}. For a pair of interacting user $u$ and item $i$, we can obtain an item $j$ not interacting with $u$ by negative sampling. We define the score of the positive instance $(u, i)$ as $\tilde y_{ui} = \mathbf{x}_u^\mathrm{T}\mathbf{x}_i$, and the score of the negative instance $(u, j)$ as $\tilde y_{uj} = \mathbf{x}_u^\mathrm{T}\mathbf{x}_j$. Then the loss function of MF is derived as follows:
\begin{equation}\label{eq1}
L(\mathbf{X}) =  - \sum\limits_{u = 1}^{\left| M \right|} {\sum\limits_{i \in {\mathcal{N}_u}} {\sum\limits_{j \notin {\mathcal{N}_u}} {ln\sigma (\tilde y_{ui} - \tilde y_{uj}}) + \lambda \left\|\mathbf{X}\right\|^2} }.
\end{equation}
where $\mathcal{N}_u$ is the set of all neighbors of the node $u$. With the loss function in Eq. (\ref{eq1}), MF tends to maximize the scores of positive instances while minimize that of negative instances, so as to integrate the historical interaction information into the user-item embedding table.

At the stage of inference, MF predicts the score of each pair of user and item, and then generates an item preference ranking list for each user relying on the scores.

\subsubsection{LightGCN}

LightGCN~\cite{he2020lightgcn}, which removes the unnecessary feature transformation and nonlinear activation operations inherited from graph convolution networks, achieves efficient and effective recommendation with its neighborhood aggregation and layer combination.

At the $k$-th layer, LightGCN obtains node embeddings $\mathbf{X}^{(k)}$ by neighborhood aggregation mechanism as follows:
\begin{equation}
       \mathbf{X}^{(k)} = \left(\mathbf{D}^{-\frac{1}{2}}\mathbf{A}\mathbf{D}^{-\frac{1}{2}}\right)\mathbf{X}^{(k-1)}.
\end{equation}
where $\mathbf{D}$ is a diagonal matrix of $\left| \mathcal{V} \right| \times \left| \mathcal{V} \right|$, $\mathbf{D}_{ii}$ indicates the number of non-zero elements in $i$-th row of $\mathbf{A}$, and $\mathbf{X}^{(0)} = \mathbf{X}$ is the user-item embedding table of the model. 
Finally, the output vectors can be achieved by combining the node representations of all layers as follows:
\begin{equation}
\mathbf{O}  =  \alpha _0\mathbf{X} + \alpha _1\mathbf{X}^{(1)} + \alpha _2\mathbf{X}^{(2)} +  \cdots  + \alpha _K\mathbf{X}^{(K)}.
\end{equation}
where the coefficients $\alpha_0$, $\alpha_1$, $\cdots$, $\alpha_K$ are conventionally $1$, and $K$ is the total layer number of LightGCN.

Following the previous studies~\cite{he2020lightgcn}, we use BPR loss function to optimize LightGCN. Let
${\mathbf{O}} = \{ {{\mathbf{o}}_1},{{\mathbf{o}}_2}, \ldots ,{{\mathbf{o}}_{M + N}}\}$ denotes the output node vectors,
we define the score of the positive instance $(u, i)$ as $\hat y_{ui} = \mathbf{o}_u^\mathrm{T}\mathbf{o}_i$, and the score of the negative instance $(u, j)$ as $\hat y_{uj} = \mathbf{o}_u^\mathrm{T}\mathbf{o}_j$. Then the loss function of LightGCN is defined as follows:
\begin{equation}
       L(\mathbf{X}, \mathbf{A}) =  - \sum\limits_{u = 1}^{\left| M \right|} {\sum\limits_{i \in {\mathcal{N}_u}} {\sum\limits_{j \notin {\mathcal{N}_u}} {ln\sigma (\hat y_{ui} - \hat y_{uj}}) + \lambda \left\|\mathbf{X}\right\|^2} }.
\end{equation}

\subsection{Lottery Ticket Hypothesis in Media Recommender Systems}\label{sec3.3}

In order to extend lottery ticket hypothesis to the field of recommendation, we explore lottery ticket hypothesis in media recommender systems (LTH-MRS) to find winning tickets in the most basic parameter of recommender models—user-item embedding tables. For a deep media recommender model, we assume its large dense user-item embedding table contains a small sparse sub-matrix, and when independently training the model with this sub-matrix, it can match or even outperform that with the original large matrix. This sub-matrix is called a winning ticket in media recommender systems.

Assume that a media recommender model $f(\mathcal{G};\mathbf{X})$ achieves the best test result of $a$ after $j$ epochs of training. Let $\mathbf{M} = \{ 0,1\}  \in \mathbb{R}^{\|\mathbf{X}\|_0}$ denotes a binary mask, then $\mathbf{M} \odot \mathbf{X}$ denotes a sparse sub-matrix of $\mathbf{X}$. Assume that $f(\mathcal{G};\mathbf{M} \odot \mathbf{X})$ achieves a test result of $a'$ after $j'$ epochs of training. Formally, LTH-MRS can be formulated as: 
$\exists {\mathbf{M}}{\text{, s}}{\text{.t}}{\text{. j'}} \leqslant {\text{j, a'}} \geqslant {\text{a and }}{\left\| {\mathbf{M}} \right\|_0} \ll {\left\| {\mathbf{X}} \right\|_0}$. The sub-matrix satisfying LTH-MRS is called a winning ticket in media recommender systems.

Once LTH-MRS is verified and the winning tickets in media recommender systems are found, the performance of a dense model can be achieved merely with far fewer parameters. The costs of memory, training and inference can be reduced effectively, providing more possibilities for the production deployment of large scale of media recommender systems.

\subsection{Identifying Winning Tickets in Media Recommender Systems}\label{sec3.4}

\begin{figure}[bt]
       \centering
       \includegraphics[scale=0.32]{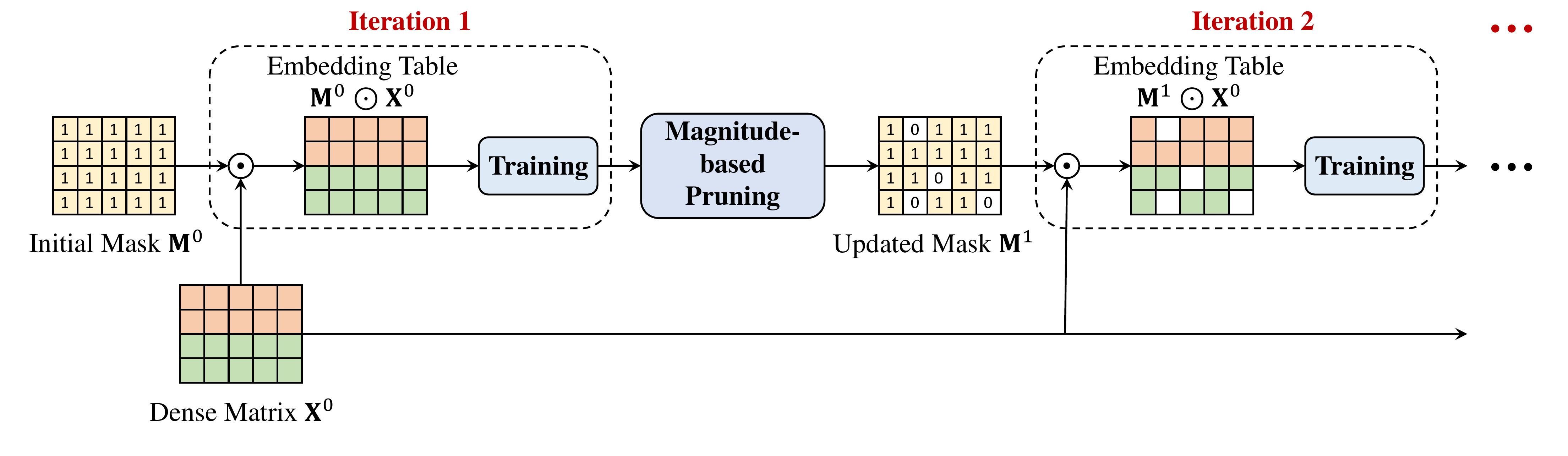}
       \caption{Overview of our used IMP algorithm to identify the winning tickets in media recommender systems.}
       \label{figure1}
\end{figure}

To verify LTH-MRS, we leverage the IMP algorithm to find the winning tickets of user-item embedding tables for media recommender models. The IMP algorithm process is shown in Figure \ref{figure1}. We use an untrainable matrix $\mathbf{M}$, where each element is a binary flag indicating whether the corresponding parameter has been pruned or not, to prevent the pruned parameters from participating in computing. Let $\mathbf{X}^0$ denotes a randomly initialized dense user-item embedding table, and $\mathbf{M}^0 = 1 \in \mathbb{R}^{\|\mathbf{X}^0\|_0}$ denotes the initial mask which is an all-ones matrix. Then the original dense model $f(\mathcal{G};\mathbf{X}^0)$ is equivalent to $f(\mathcal{G};\mathbf{M}^0 \odot \mathbf{X}^0)$. We perform magnitude-based sparse pruning on the embedding table $\mathbf{M}^0 \odot \mathbf{X}^0$ by setting the values at the corresponding positions in $\mathbf{M}^0$ to $0$. We define the percentage of parameters to be pruned at each iteration as the iterative pruning rate $pr\%$, and the proportion of 0 in $\mathbf{M}^i$ (i.e. the mask after $i$ pruning iterations) as the sparsity $\frac{\left\| {\mathbf{M}^i} \right\|_0}{\left\| {\mathbf{M}^0} \right\|_0}$. Then, an embedding sub-matrix $\mathbf{M}^I \odot \mathbf{X}^0$ with the sparsity of $1-(1-pr\%)^I$ can be achieved after $I$ pruning iterations.

To find the winning tickets, we run the following operations at $i$-th iteration:

(1) \textbf{Training}: train the model $f(\mathcal{G};\mathbf{M}^i \odot \mathbf{X}^0)$ for $J$ epochs to achieve $f(\mathcal{G};\mathbf{M}^i \odot \mathbf{X}^J)$.

(2) \textbf{Pruning}: prune $pr\%$ of non-zero lowest magnitude values in $f(\mathcal{G};\mathbf{M}^i \odot \mathbf{X}^J)$ to generate a updated mask $\mathbf{M}^{i+1}$.

(3) \textbf{Rewinding}: reset the updated dense matrix $\mathbf{X}^J$ to $\mathbf{X}^0$, creating a new lottery ticket $\mathbf{M}^{i+1} \odot \mathbf{X}^0$.

Note that the operation (3) is called rewinding mechanism, which is proposed by Frankle and Carbin~\cite{frankle2018lottery}. They believed that only when the small sparse sub-networks obtained the same initialization of the original large dense networks, can they be trained independently and efficiently.

\begin{algorithm}[bt]
       \caption{An IMP algorithm for the user-item embedding table.}\label{alg1}
      \KwIn{A user-item bipartite graph ${\cal G}$, a recommender model $f(\mathcal{G};\mathbf{X}^0)$, the iterative pruning rate $pr\%$, the total pruning iteration $I$ and the training epoch $J$.
      }
      \KwOut{The lottery ticket set $\mathcal{S}$.}
      Initialize a mask $\mathbf{M}^0 \leftarrow 1 \in \mathbb{R}^{\left\| {\mathbf{X}^0} \right\|_0}$.
      
      Initialize a lottery ticket set $\mathcal{S} \leftarrow \emptyset$.

       \For{$i \leftarrow 0$ to $I-1$}
       {
           \textbf{Training}: train $f(\mathcal{G};\mathbf{M}^i \odot \mathbf{X}^0)$ with ${\cal G}$ for $J$ epochs to achieve $f(\mathcal{G};\mathbf{M}^i \odot \mathbf{X}^J)$.

           \textbf{Pruning}: select $pr\%$ of non-zero lowest magnitude values in $\mathbf{M}^i \odot \mathbf{X}^J$, and set the values at the corresponding positions in $\mathbf{M}^i$ to $0$ to generate $\mathbf{M}^{i+1}$.

           \textbf{Rewinding}: reset $\mathbf{X}^J$ to $\mathbf{X}^0$ to generate a new lottery ticket $\mathbf{M}^{i+1} \odot \mathbf{X}^0$.

           $\mathcal{S} \leftarrow \mathcal{S} \cup \left\{ \mathbf{M}^{i+1} \odot \mathbf{X}^0 \right\}$
       }
\end{algorithm}

The specifical process is detailed in Algorithm \ref{alg1}. When Algorithm \ref{alg1} is completed (i.e., after $I$ iterations of pruning), we can achieve a lottery ticket set $\mathcal{S} = \left\{ \mathbf{M}^{1} \odot \mathbf{X}^0, \mathbf{M}^{2} \odot \mathbf{X}^0, \cdots, \mathbf{M}^{I} \odot \mathbf{X}^0 \right\}$, where each ticket is a sparse embedding sub-matrix with the sparsity of $1-(1-pr\%), 1-(1-pr\%)^2, \cdots, 1-(1-pr\%)^I$, respectively. Finally, we independently train and test each lottery ticket in $\mathcal{S}$ to verify the existence of winning tickets.

\subsection{Complexity Analysis}\label{sec3.5}

Here, we provide the complexity analyses for MF and LightGCN to show the superiority of the winning ticket.

\textbf{MF}. The inference time complexity of the original dense MF is $\mathcal{O}\left( M \times N \times F \right)$ and the memory complexity is $\mathcal{O}\left( M \times F + N \times F \right)$, where $M$ and $N$ are the number of users and items in the recommender system, and $F$ is the size of user-item embedding vector. For a winning ticket $\mathbf{M} \odot \mathbf{X}$ of MF, we denote $\mathbf{M}_u$ as the mask of the user $u \in \mathcal{U}$, and $\mathbf{M}_i$ as the mask of  the item $i \in \mathcal{I}$, where $\mathcal{U}$ and $\mathcal{I}$ are the sets of user and item, respectively. Then the inference time complexity of the winning ticket is $\mathcal{O}\left( M \times N \times \min \left( {F_u^*,F_i^*} \right) \right)$, and the memory complexity is $\mathcal{O}\left( M \times F_u^* + N \times F_i^* \right)$, where $F_u^* = \max \limits_{u \in \mathcal{U}} \left\|\mathbf{M}_u \right\|_0$ indicates the maximum size of sparse user embeddings, and $F_i^* = \max \limits_{i \in \mathcal{I}} \left\|\mathbf{M}_i \right\|_0$ indicates the maximum size of sparse item embeddings.

\textbf{LightGCN.} For a $K$-layer LightGCN, the time complexity of neighborhood aggregation is $\mathcal{O}\left( K \times \|\mathbf{A}\|_0 \times F \right)$, where $\mathbf{A}$ is the adjacency matrix, the time complexity of layer combination is $\mathcal{O}\left( M \times F + N \times F \right)$ and the time complexity of similarity computation is $\mathcal{O}\left( M \times N \times F \right)$. Considering $M + N \ll M \times N$, the inference time complexity of the original dense LightGCN is $\mathcal{O}\left( K \times \|\mathbf{A}\|_0 \times F + M \times N \times F \right)$, and the memory complexity is $\mathcal{O}\left( M \times F + N \times F \right)$. For a winning ticket of LightGCN, the inference time complexity is $\mathcal{O}\left( K \times \|\mathbf{A}\|_0 \times \min \left( {F_u^*,F_i^*} \right) + M \times N \times \min \left( {F_u^*,F_i^*} \right) \right)$, and the memory complexity is $\mathcal{O}\left( M \times F_u^* + N \times F_i^* \right)$.

As compared with the large dense models, the winning tickets represent users and items with much smaller sparse vectors (i.e., $F_u^*,F_i^* \ll F$). Therefore, the inference time complexity and the memory complexity of the winning tickets are far lower than that of the original dense models.

\section{Experiments}
Beginning with an introduction of the experiment settings (\ref{sec4.1}), this section mainly attempts to explore the following questions through a series of comparison experiments:

Q1: Does the winning ticket in media recommender systems exists? (\ref{sec4.2})

Q2: Does the IMP algorithm reliably find winning tickets? (\ref{sec4.3})

Q3: Does the found winning tickets outperform the other compression models? (\ref{sec4.4})

To further explore the effects of different implementation details of IMP, we conduct ablation studies (\ref{sec4.5}).
On top of that, we also compare the training speed of the winning tickets and the original embedding table (\ref{sec4.6}).
Finally, we visualize the winning tickets for interpretable analysis (\ref{sec4.7}).

\subsection{Experiment Settings}\label{sec4.1}
Here, we will detail our experiment settings, including the datasets, the evaluation metrics, the backbone models and the hyperparameters settings.

\textbf{Datasets.} We conducted extensive experiments on three public benchmark datasets:

\begin{itemize}
       \item \textbf{Yelp2018}~\endnote{https://www.kaggle.com/yelp-dataset/yelp-dataset} is a dataset for recommending restaurants and bars, presented in the 2018 Yelp Dataset Challenge.
       \item \textbf{TikTok}~\endnote{https://www.biendata.xyz/competition/icmechallenge2019} is a dataset for recommending the TikTok videos, presented in the 2019 ICME Challenge.
       \item \textbf{Kwai}~\endnote{https://www.kuaishou.com/activity/uimc} is a dataset for recommending the Kwai videos, presented in the China MM 2018.
\end{itemize}

The TikTok and Kwai datasets are both the real-world user-video interaction datasets under the media convergence environment. Following the previous works~\cite{wang2019neural,he2017neural}, we split the datasets into training set, validation set and test set with a ratio of 7:1:2. Table \ref{table1} shows overall statistics for the three datasets. Note that the original interaction data of TikTok is too huge, so we merely use the videos within a period, and filter the data without complete modalities.

\begin{table}[bt]
\small
\caption{Statistics of the datasets.}\label{table1}
\begin{threeparttable}
\begin{tabular}{ccccc}
\headrow
\thead{Datasets} & \thead{User Number} & \thead{Items Number} & \thead{Interaction Number} & \thead{Density}\\
Yelp2018 & 31,668 & 38,048 & 1,561,406 & 0.00130\\
TikTok & 36,638 & 97,117 & 746,546 & 0.00021\\
Kwai & 7,010 & 80,631 & 292,042 & 0.00052\\
\hline  
\end{tabular}
\end{threeparttable}
\end{table}

\textbf{Evaluation Metrics.} We adopt Recall and Normalized Discounted Cumulative Gain (NDCG) as the main experimental evaluation metrics. For users in the test set, we follow the all-ranking protocol to evaluate the top-$K$ recommendation performance and report the average Recall@K and NDCG@K, where we set $K=20$.

\textbf{Backbone Models.}
We adopt MF and LightGCN as our backbone models. MF and LightGCN are representative because they are both merely user-item interaction-based deep recommender models without any external information, and the user-item embedding table is the only model parameter. (See \ref{sec3.2} for more details)

\textbf{Hyper-parameters Settings.}
We apply the Adam~\cite{kingma2014adam} as the optimizer with the learning rate of 0.001, the L2-regularization weight $\lambda$ of 0.0001 and the batch size of 2048. In main experiments, the training epoch number $J$ is 1000, the iterative pruning rate $pr\%$ is 0.1 and the pruning iteration number $I$ is 30.

\begin{table}[b]
       \small
      \caption{The Recall@20 and NDCG@20 of the sub-models with different sparsity levels of MF and LightGCN on Yelp2018, TikTok and Kwai datasets. The bold results indicate the performance of winning tickets, and the sparsity of 0 indicates the original dense model.}\label{table2}
      \begin{threeparttable}
      \begin{tabular}{c|c|cccc|cccc}
      \headrow
      \multicolumn{2}{c|}{\textbf{Evaluation Metrics}} & \multicolumn{4}{c|}{\textbf{Recall@20}} & \multicolumn{4}{c}{\textbf{NDCG@20}}\\
      \multicolumn{2}{c|}{Sparsity (\%)} & 0 & 27.1 & 65.13 & 95.29 & 0 & 27.1 & 65.13 & 95.29 \\
      \hline
      \hiderowcolors
      \multirow{2}*{Yelp2018} & MF & 0.0459 & \textbf{0.0487} & \textbf{0.0468} & 0.0395 & 0.0372 & \textbf{0.0396} & \textbf{0.0378} & 0.0320\\
      ~ & LightGCN & 0.0600 & \textbf{0.0602} & 0.0584 & 0.0510 & 0.0488 & \textbf{0.0491} & 0.0476 & 0.0418\\
      \hline
      \multirow{2}*{TikTok} & MF & 0.0851 & \textbf{0.0955} & \textbf{0.0961} & 0.0638 & 0.0500 & \textbf{0.0552} & \textbf{0.0557} & 0.0354\\
      ~ & LightGCN & 0.1423 & \textbf{0.1539} & \textbf{0.1577} & 0.1379 & 0.0828 & \textbf{0.0898} & \textbf{0.0918} & 0.0813\\
      \hline
      \multirow{2}*{Kwai} & MF & 0.0411 & \textbf{0.0464} & \textbf{0.0525} & \textbf{0.0529} & 0.0318 & \textbf{0.0353} & \textbf{0.0393} & \textbf{0.0431}\\
      ~ & LightGCN & 0.0779 & \textbf{0.0847} & \textbf{0.0857} & 0.0751 & 0.0638 & 0.0679 & \textbf{0.0689} & 0.0614\\
      \hline
      \end{tabular}
      \end{threeparttable}
\end{table}

\subsection{Existence of the Winning Ticket}\label{sec4.2}

We apply the IMP algorithm on Yelp2018, TikTok and Kwai datasets to obtain a series of sparse sub-networks (a.k.a. lottery tickets) of MF and LightGCN. Then each lottery ticket will be trained and tested independently to determine whether it wins or not. We present the experimental results of the 32-dimensional original embedding size in Table \ref{table2}. As shown in these results, we have found a lot of winning tickets, which can achieve significantly better performance with far fewer parameters than the original models. Taking the results on Kwai dataset for example, as compared with the original MF, the winning ticket improves Recall@20 and NDCG@20 by 28.7\% and 35.5\%, respectively, while reducing 95.29\% parameters; as compared with the original LightGCN, the winning ticket improves Recall@20 and NDCG@20 by 10.0\% and 7.99\%, respectively, while reducing 65.13\% parameters.
The above observations show that LTH-MRS is valid, and the winning tickets exist widely in media recommender systems.

\subsection{Effectiveness of the IMP Algorithm}\label{sec4.3}

\begin{figure}[bt]
       \centering
       \includegraphics[scale=0.35]{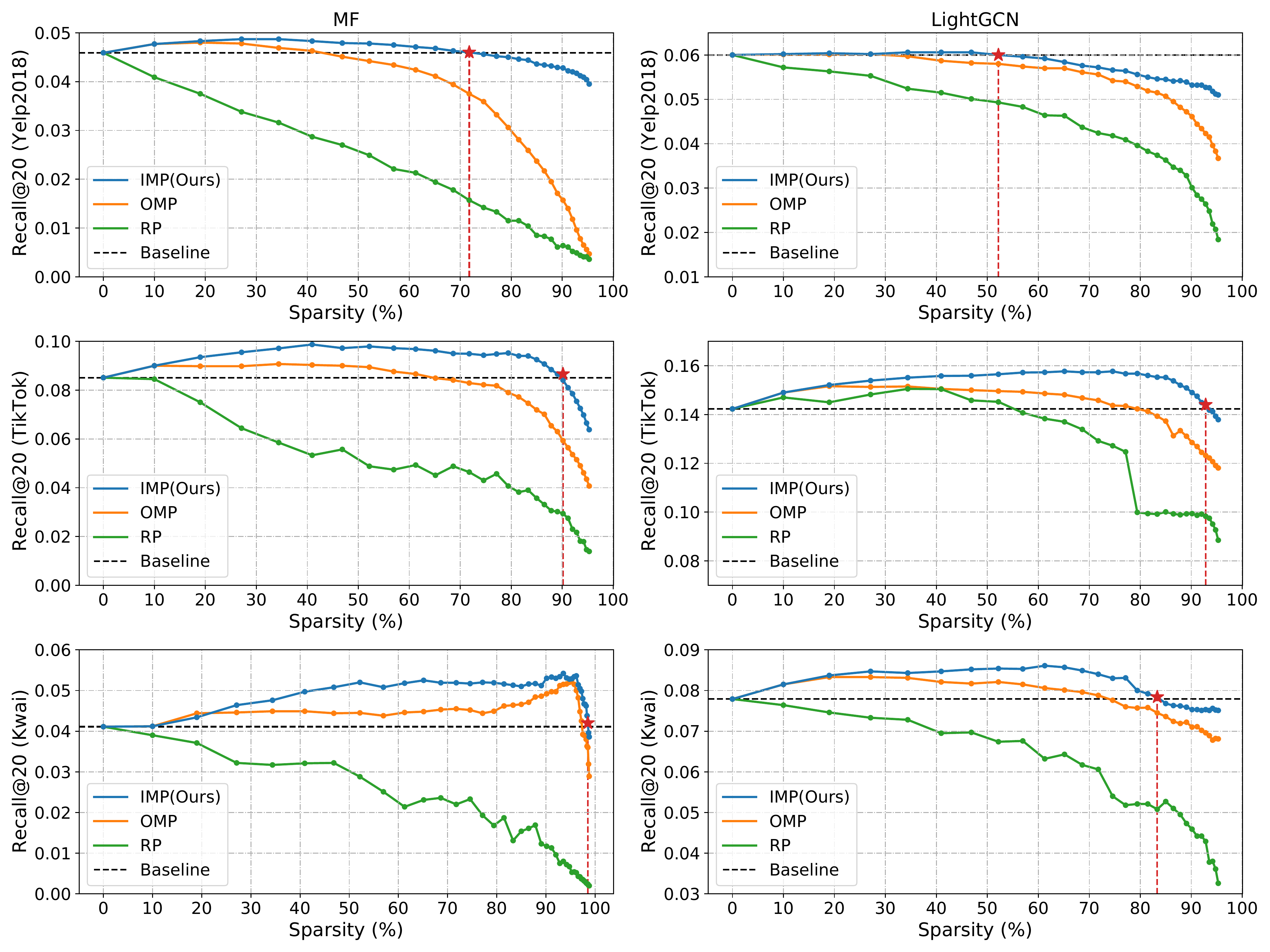}
       \caption{The Recall@20 results of the lottery tickets with different sparsity of MF and LightGCN obtained by IMP, RP and OMP approaches on Yelp2018, TikTok and Kwai datasets. The positions of \textit{red stars} ($\color{symbol-red} \star$) indicate the highest sparsity that winning tickets can achieve while preserving comparable performance.}
       \label{figure2}
\end{figure}

To explore whether the IMP algorithm can find the winning tickets stably and effectively, we select two representative pruning approaches\cite{frankle2018lottery} for comparison: \textbf{Random Pruning (RP)} randomly removes elements from a trained embedding table, and \textbf{One-shot Magnitude-based Pruning (OMP)} directly reduce a trained embedding table to the target sparsity based on the element magnitude without iterations. 
We compare the performance of the lottery tickets obtained by IMP, RP and OMP, with the setting of 32-dimensional original embedding size, to demonstrate the effectiveness of the IMP algorithm for finding winning tickets. Notably, the \textit{Baselines} with black dotted lines in Figure 2-5 indicate the corresponding large-dense recommender models. 

From the results shown in Figure \ref{figure2}, we can observe that:

(1) RP can not stably find the winning tickets. In our experiments, RP merely found winning tickets of LightGCN on TikTok dataset. Moreover, the winning tickets found by RP achieved 40.7\% and 27.3\% declines in the highest sparsity, and achieved 4.8\% and 0.7\% declines in highest Recall@20, compared with that by IMP and OMP, respectively.

(2) IMP and OMP can both stably find the winning tickets, but the winning tickets found by IMP significantly and consistently outperform that by OMP. Specifically, on Yelp2018, TikTok and Kwai datasets, as compared with OMP, the winning tickets of MF found by IMP achieve 30.8\%, 25.2\% and 2.5\% improvements in the highest sparsity, and achieve 18.8\%, 8.8\% and 1.1\% improvements in the highest Recall@20; the winning tickets of LightGCN found by IMP achieve 22.0\%, 15.6\% and 8.9\% improvements in the highest sparsity, and achieve 6.6\%, 4.0\% and 3.4\% improvements in the highest Recall@20.

The above observations show that our used IMP algorithm can reliably find the winning tickets, and the winning tickets found by IMP always outperform that by the compared approaches.

\subsection{Performance of the Winning Ticket}\label{sec4.4}

To further demonstrate the performance of the found winning tickets, we compare against two competitive baselines: \textbf{Linear Compression Model (LCM)}~\endnote{https://github.com/gusye1234/KD\_on\_Ranking}, which compresses the large embedding table into a small \textit{dense} matrix with trainable feature transformation; and \textbf{Plug-in Embedding Pruning model (PEP)}~\cite{liu2020learnable}, which prunes the embedding table into a small \textit{sparse} matrix with a trainable dynamic pruning threshold.

\begin{table}[bt]
        \small
       \caption{The Recall@20 and NDCG@20 of the winning tickets, LCM and PEP with different sparsity levels of LightGCN on Yelp2018, TikTok and Kwai datasets.}\label{table3}
       \begin{threeparttable}
       \begin{tabular}{c|c|cc|cc|cc|cc}
       \headrow
       \multicolumn{2}{c|}{\textbf{Evaluation Metrics}} & \multicolumn{4}{c|}{\textbf{Recall@20}} & \multicolumn{4}{c}{\textbf{NDCG@20}}\\
       \multicolumn{2}{c|}{Original Size} & \multicolumn{2}{c|}{64} & \multicolumn{2}{c|}{128} & \multicolumn{2}{c|}{64} & \multicolumn{2}{c}{128} \\
       \multicolumn{2}{c|}{Sparsity (\%)} & 50 & 75 & 50 & 75 & 50 & 75 & 50 & 75 \\
       \hline
       \hiderowcolors
       \multirow{3}*{Yelp2018} & LCM & 0.0558 & 0.0459 & 0.0610 & 0.0537 & 0.0460 & 0.0374 & 0.0493 & 0.0438\\
       ~ & PEP & 0.0584 & 0.0570 & 0.0627 & 0.0601 & 0.0473 & 0.0459 & 0.0507 & 0.0483\\
       ~ & Winning Ticket & \textbf{0.0635} & \textbf{0.0609} & \textbf{0.0666} & \textbf{0.0651} & \textbf{0.0525} & \textbf{0.0500} & \textbf{0.0545} & \textbf{0.0530}\\
       \hline
       \multirow{3}*{TikTok} & LCM & 0.1508 & 0.1341 & 0.1594 & 0.1500 & 0.0895 & 0.0791 & 0.0941 & 0.0885\\
       ~ & PEP & 0.1640 & 0.1567 & 0.1814 & 0.1801 & 0.0956 & 0.0901 & 0.1068 & 0.1047\\
       ~ & Winning Ticket & \textbf{0.1851} & \textbf{0.1829} & \textbf{0.1911} & \textbf{0.1913} & \textbf{0.1085} & \textbf{0.1064} & \textbf{0.1128} & \textbf{0.1130}\\
       \hline
       \multirow{3}*{Kwai} & LCM & 0.0690 & 0.0640 & 0.0742 & 0.0652 & 0.0585 & 0.0542 & 0.0544 & 0.0538\\
       ~ & PEP & 0.0840 & 0.0836 & 0.0862 & 0.0860 & 0.0684 & 0.0673 & 0.0696 & 0.0695\\
       ~ & Winning Ticket & \textbf{0.0867} & \textbf{0.0856} & \textbf{0.0888} & \textbf{0.0882} & \textbf{0.0696} & \textbf{0.0693} & \textbf{0.0706} & \textbf{0.0705}\\
       \hline
       \end{tabular}
       \end{threeparttable}
\end{table}

We present the comparison results of winning tickets, LCM and PEP for LightGCN in Table \ref{table3}, demonstrating that our found winning tickets remarkably outperform LCM and PEP with equivalent quantities of parameters. Taking the results on Yelp2018 dataset as examples, for a $128$-dimensional dense embedding table, as compared with LCM, the winning tickets improve Recall@20 and NDCG@20 by 9.18\% and 10.54\% when the sparsity is 50\%, and improves that by 21.23\% and 21.00\% when the sparsity is 25\%.

As aforementioned, it is generally considered that dense models are easier to train for better performance than sparse models. However, the experimental results show that the \textit{sparse} compressed models—winning tickets and PEP, can consistently outperform the \textit{dense} compressed model—LCM, which seems a little counter-intuitive. The reason we believed is that, as a dense model, LCM represents different users and items with a uniform size, neglecting the diversity and the specificity among users and items. Hence, it may be hard for LCM to handle the heterogeneity of users and items with different popularities. By contrast, our used IMP algorithm automatically determines an appropriate embedding size for each user or item by pruning unimportant parameters during searching for winning tickets, and PEP also achieves this property with its dynamic threshold mechanism. As compared with LCM, winning tickets and PEP can express the heterogeneity of users and items better by assigning different embedding sizes for them. The various sparsity of embedding vectors describes the intrinsic diversity among different users and items, and thus improves the performance of winning tickets.

Moreover, as shown in Table 3, although the dynamic threshold mechanism in PEP remarkably speeds up the process of pruning, the performance of final sparse models is worse than that of winning tickets. In addition, the performance of PEP highly depends on hyperparameters for the trainable pruning threshold, so it is required to consume expensive time for tuning those hyperparameters carefully. In terms of performance, simplicity and stability, the winning tickets has shown great superiority as compared with PEP.

\begin{figure}[bt]
       \centering
       \includegraphics[scale=0.35]{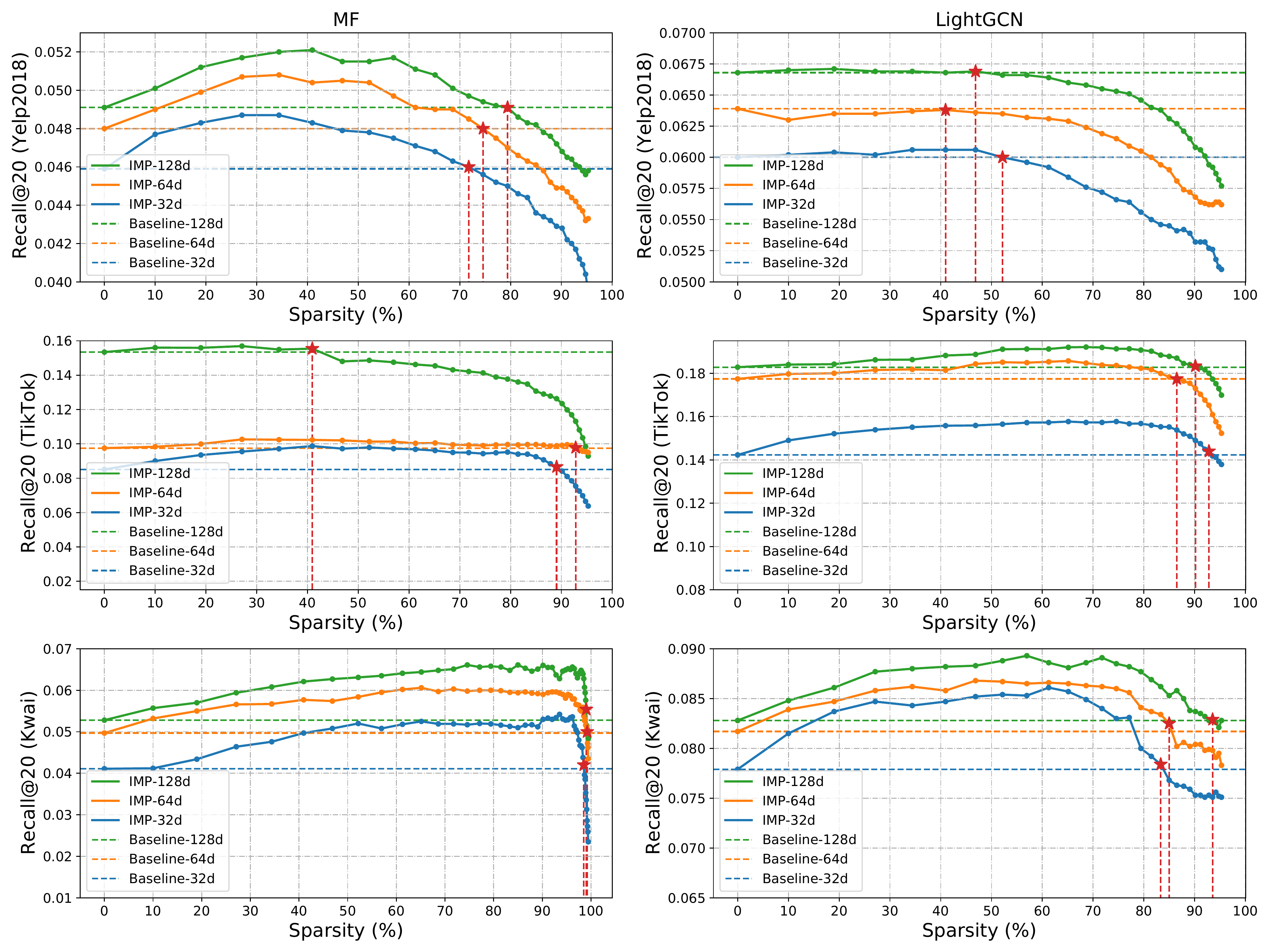}
       \caption{The Recall@20 results of the lottery tickets of MF and LightGCN with different original embedding sizes on Yelp2018, TikTok and Kwai datasets, where the \textit{IMP-$i$d} indicates the IMP with original embedding size of $i$.}
       \label{figure3}
\end{figure}

\begin{figure}[!h]
       \centering
       \includegraphics[scale=0.47]{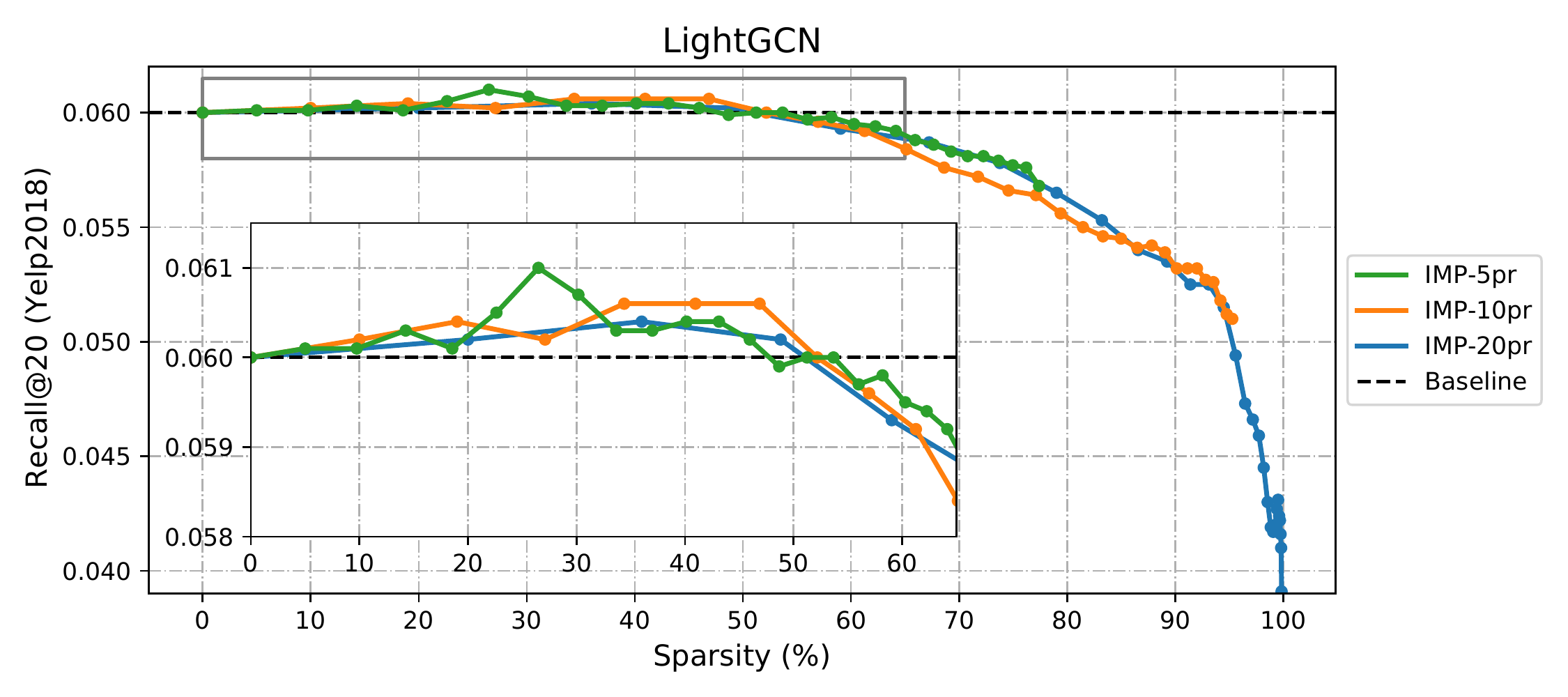}
       \caption{The performance comparison of IMP algorithm with different iterative pruning rates, where the \textit{IMP-$i$pr} indicates the IMP with iterative pruning rate of $i$.}
       \label{figure4}
\end{figure}

\subsection{Ablation Study}\label{sec4.5}

To further explore the effects and contributions of different implementation details of IMP, including the original embedding size, the iterative pruning rate $pr\%$ and the rewinding mechanism, we conduct a series of ablation studies.

\textbf{Effect of Original Embedding Size.}
We present the performance of winning tickets with different original embedding sizes in Figure \ref{figure3}, from which it can be observed that: (1) for the same deep recommender models with different sizes, with the increase of sparsity, the performance trends of lottery tickets are almost consistent; (2) for deep recommender models with different original embedding sizes, the winning tickets widely exist and always can be found by our used IMP algorithm.

\textbf{Effect of Iterative Pruning Rate.}
We utilize IMP algorithm with different $pr\%$ (5\%, 10\% and 20\%) to find winning tickets of LightGCN on Yelp2018 dataset, and present the results in Figure \ref{figure4}, from which it can be observed that: (1) the highest sparsity achieved with different $pr\%$ is almost the same; (2) lower $pr\%$ achieves higher best Recall@20; (3) higher $pr\%$ takes fewer steps to reach the sparsest winning ticket. These observations demonstrate that a high iterative pruning rate can effectively improve the search efficiency, while it may miss the optimal winning ticket. Therefore, to trade off effectiveness and efficiency, it is necessary to set a reasonable $pr\%$ when applying the IMP algorithm to find winning tickets.

\begin{figure}[bt]
       \centering
       \includegraphics[scale=0.35]{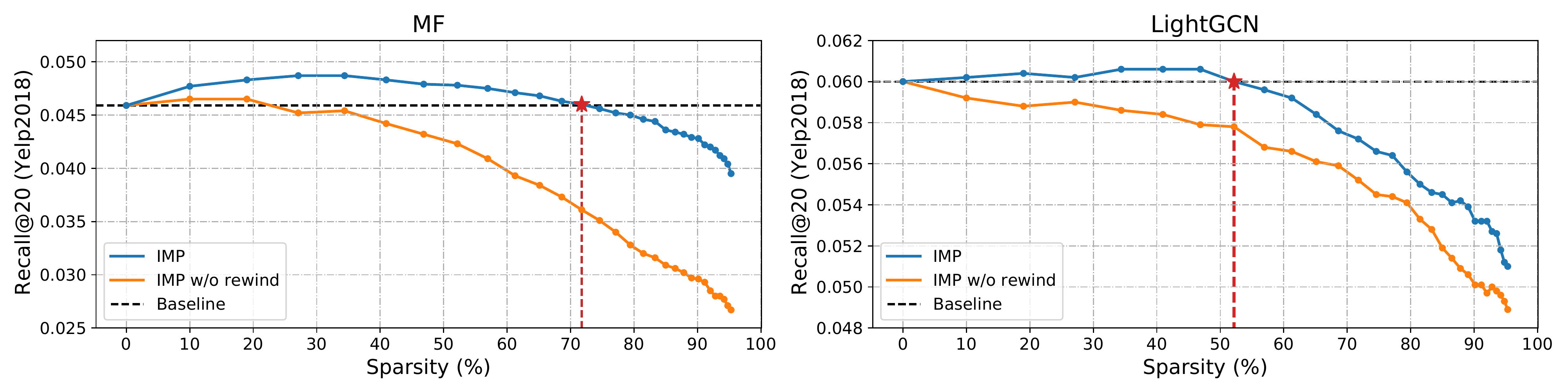}
       \caption{The ablation study about rewinding mechanism, where the \textit{IMP w/o rewind} indicates the IMP algorithm without rewinding mechanism.}
       \label{figure5}
\end{figure}

\textbf{Effect of Rewinding Mechanism.}
To explore the importance of rewinding mechanism in our IMP algorithm, we conduct ablation studies for MF and LightGCN on Yelp2018 dataset. It can be observed in Figure \ref{figure5} that compared with \textit{IMP w/o rewind}, (1) the IMP with rewinding mechanism can find winning tickets more stably and reliably; (2) the IMP with rewinding mechanism improves the performance in Recall@20 and sparsity significantly and consistently.

Overall, in IMP algorithm, the iterative pruning is a search technique for winning tickets, while the rewinding mechanism ensures the stability of the search and the effectiveness of the winning tickets.

\begin{figure}[bt]
       \centering
       \includegraphics[scale=0.50]{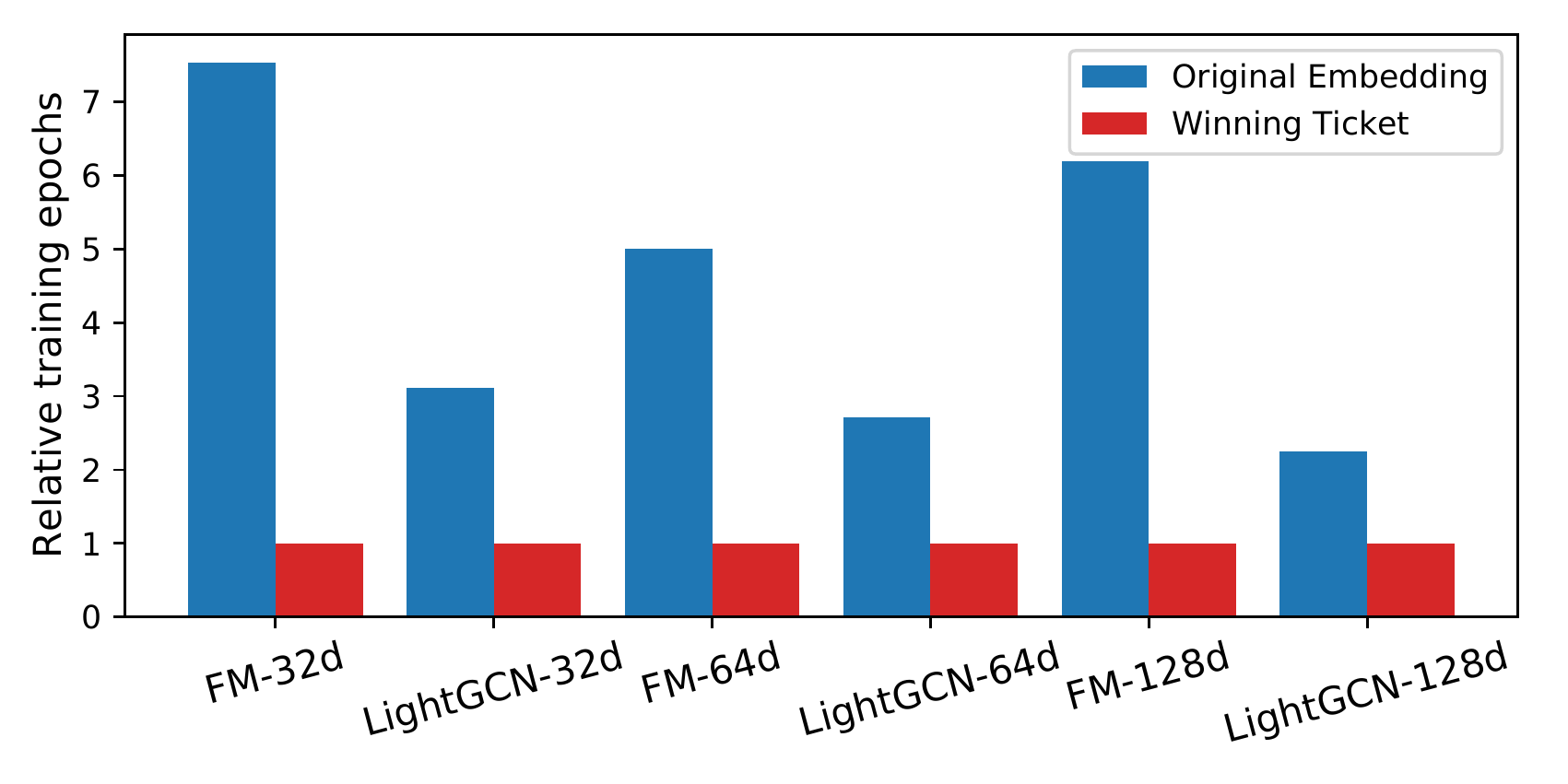}
       \caption{Relative training epochs of winning tickets and original embeddings.}
       \label{figure6}
\end{figure}

\subsection{Comparison of Training Speed}\label{sec4.6}
We compare the training speed of winning tickets and original embeddings to show that during independently training, the winning tickets can achieve comparable performance to the original dense embedding table with faster learning speed. We take the consumed training epochs of winning tickets as one unit, and then present the relative training epochs of original embeddings in Figure \ref{figure6}, from which we can observe that the training speed of winning tickets is far higher than that of original embeddings. For example, the winning ticket of MF-32d reached comparable performance with a learning speed 8 times the original embedding table. These experimental results demonstrate the huge potential of the winning ticket for reducing expensive training cost of large-scale media recommender systems.

\subsection{Case Study}\label{sec4.7}
\begin{figure}[bt]
       \centering
       \includegraphics[scale=0.27]{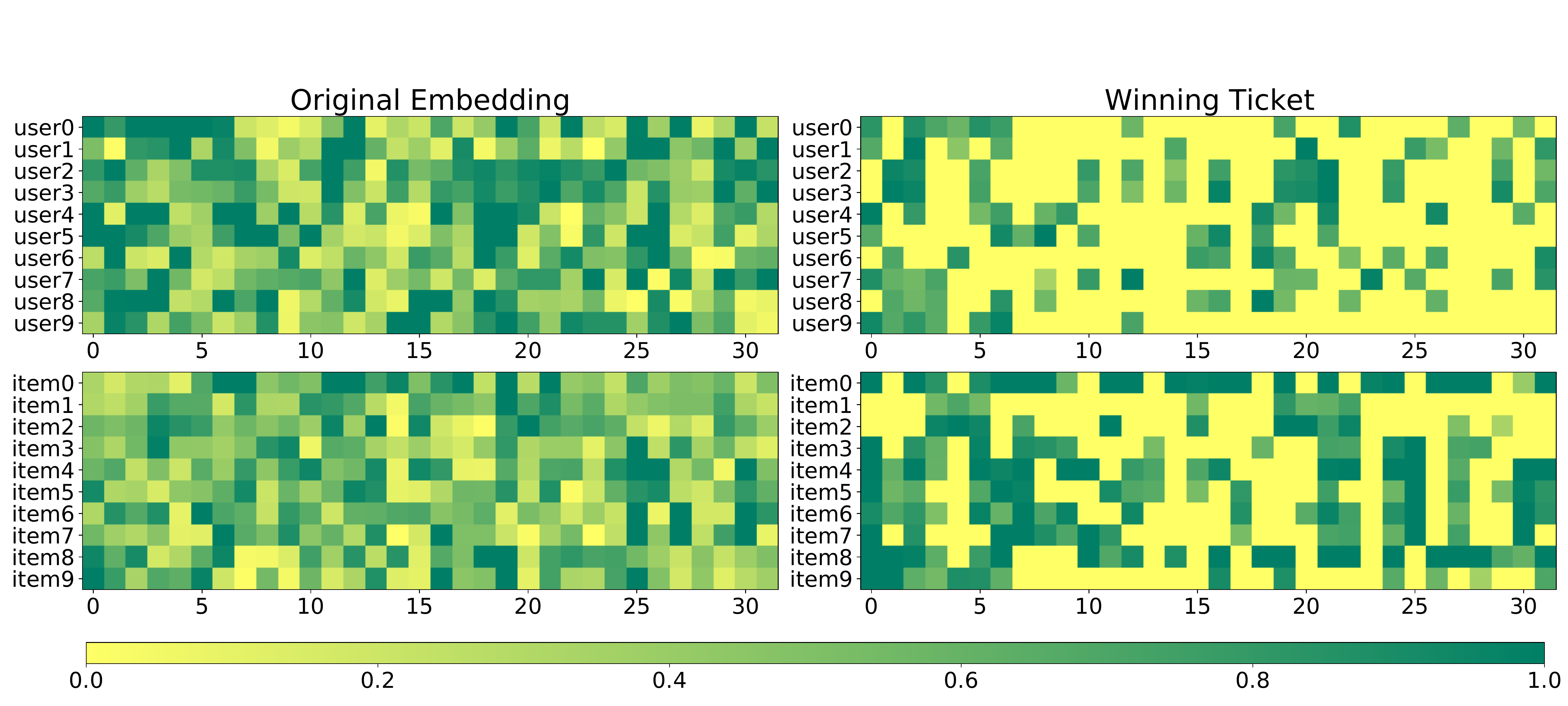}
       \caption{Visualization of winning ticket and original embedding.}
       \label{figure7}
\end{figure}

We visualize the original embedding table and the winning ticket on Yelp2018 dataset for case study. Due to the limited space, we merely show the first 10 users and items. As shown in Figure \ref{figure7}, the highly sparse winning ticket achieved by IMP tends to preserve more critical features (of high magnitude) for different users and items. During the iterations of identifying winning tickets, the IMP algorithm can reduce the parameter scale of redundancy features while preserving the significance of those meaningful features. By denoising those unimportant features, we can obtain a sparse embedding table (winning ticket) that can achieve comparable test performance of the full embedding with much fewer parameters.

\section{Conclusion and Futute Work}
To address the problem of parameter redundancy in media recommender systems, we extended the lottery ticket hypothesis to the field of recommendation. We studied lottery ticket hypothesis in media recommender systems, exploiting IMP algorithm to find winning tickets of the user-item embedding table in two representative deep recommender models --- MF and LightGCN. Empirical results on three real-world datasets showed the winning tickets can achieve better performance with much fewer parameters and faster training speed than the full user-item embeddings.

In the future, we plan to explore: (1) \textbf{efficient schemes for identifying winning tickets.} The iterative identification of IMP requires a costly train-prune-retrain process, which limits the practical benefits; (2) \textbf{winning tickets for implicit feedback interaction records}. There exists a lot of noisy data in the implicit feedback interaction records of recommender systems. For example, a noisy interaction will be produced when a user accidentally clicks on an uninterested item. Huge number of interaction records will lead to an expensive training cost, while the serious noisy interactions will mislead the learning of recommender models. (3)\textbf{inductive mode for new-coming users and items}. The proposed method of this paper is limited to the transductive setting. In each iteration, we performed pruning on the whole embedding matrix, which allowed us to optimize the networks globally, but also limited the generalization to new-coming users/items. In the future, we will consider extending our method to the inductive setting.

\section*{ACKNOWLEDGEMENTS}
This work is supported by the National Natural Science Foundation of China (U19A2079, 61972372, 62121002) and National Key Research and Development Program of China (2020YFB1406703).

\printendnotes

\bibliography{ref}

\end{document}